# Magnetic and Metal-Insulator Transitions in $\beta$-Na$_{0.5}$CoO$_2$ and $\gamma$-K$_{0.5}$CoO$_2$ -NMR and Neutron Diffraction Studies-

H. Watanabe[1], Y. Mori[1], M. Yokoi[1], T. Moyoshi[1], M. Soda[1], Y. Yasui[1], Y. Kobayashi[1], M. Sato[1*],
N. Igawa[2] and K. Kakurai[2]

[1]*Department of Physics, Division of Material Science, Nagoya University, Furo-cho, Chikusa-ku, Nagoya 464-8602*
[2] *Japan Atomic Energy Agency, Quantum Beam Science Directorate, Tokai-mura, Naka-gun, Ibaraki 319-1195*

Co-oxides $\beta$-Na$_{0.5}$CoO$_2$ and $\gamma$-K$_{0.5}$CoO$_2$ have been prepared by the Na de-intercalation from $\alpha$-NaCoO$_2$ and by the floating-zone method, respectively. It has been found that successive phase transitions take place at temperatures $T_{c1}$ and $T_{c2}$ in both systems. The appearance of the internal magnetic field at $T_{c1}$ with decreasing temperature $T$ indicates that the antiferromagnetic order exists at $T \leq T_{c1}$, as in $\gamma$-Na$_{0.5}$CoO$_2$. For $\beta$-Na$_{0.5}$CoO$_2$, the transition temperatures and the NMR parameters determined from the data taken for magnetically ordered state are similar to those of $\gamma$-Na$_{0.5}$CoO$_2$, indicating that the difference of the stacking ways of the CoO$_2$ layers between these systems do not significantly affect their physical properties. For $\gamma$-K$_{0.5}$CoO$_2$, the quantitative difference of the physical quantities are found from those of $\beta$- and $\gamma$-Na$_{0.5}$CoO$_2$. The difference between the values of $T_{ci}$ ($i$ = 1 and 2) of these systems might be explained by considering the distance between CoO$_2$ layers.

## 1. Introduction

Since the discovery of High-$T_c$ cuprates, a considerable number of studies have been made on 3$d$ transition-metal oxides with layered structures. Alkali-metal cobalt oxides A$_x$CoO$_2$ (A = alkali metal) are among such kinds of materials.[1-5] The system consists of layers of edge-sharing CoO$_6$ octahedra and alkali metals, which stack along the $c$-axis alternately. Because the layers have the triangular lattice of Co ions, the possibility of geometrical frustration is expected. Another interesting viewpoint on these materials is, as can be found in various Co-oxides, the spin state change of Co$^{3+}$ and/or Co$^{4+}$ ions.[6-8]

For Na$_x$CoO$_2$, there exist several polymorphs, $\gamma$ (0.3 < $x$ < 0.74, P2-type), $\alpha$ (0.9 < $x$ < 1.0, O3-type), $\alpha$' ($x$ = 0.75, O'3-type), and $\beta$ (0.55 < $x$ < 0.6, P'3-type) phase-ones.[9-10] Such polymorphs are classified by the geometrical coordination of the oxygen ions surrounding Na atoms (octahedral (O) coordination and trigonal prismatic (P) one) and the number of repeating unit of CoO$_2$ layers along the $c$-axis. (In the abbreviation of the structural type, ' denotes a monoclinic distortion of the unit cell.) Each structure of the polymorphs is characterized by its own stacking sequence of CoO$_2$ layers along the $c$-axis. So far, $\gamma$-Na$_x$CoO$_2$ has been investigated primarily due to its peculiarity of large thermoelectric power coexisting with low electric resistivity.[11] Physical properties of the $\gamma$-Na$_x$CoO$_2$ depend on the Na content $x$ and temperature $T$.[12] For $x$ > 0.6, the existence of strong electronic correlation is indicated by the Curie-Weiss like magnetic susceptibility $\chi$ and the electrical specific heat coefficient $\gamma$, while for $x$ < 0.6, this material exhibits the lower resistivity and pseudogap-like behavior of the magnetic susceptibility $\chi$.[12,13] Furthermore, at $x$ = 0.3, superconducting phase appears when H$_2$O molecules are intercalated between CoO$_2$ and Na layers.[14] Neutron scattering studies carried out for $\gamma$-Na$_{0.82}$CoO$_2$[15] and $\gamma$-Na$_{0.75}$CoO$_2$[16] have revealed that the interlayer magnetic interaction of the electron system is so significant that the system has the three-dimensional nature, contrary to the structural two-dimensionality. It is also reported that the in-plane magnetic correlation is ferromagnetic in these Na-rich systems. However, various kinds of results obtained by NMR/NQR, neutron scattering and other studies on $\gamma$-Na$_x$CoO$_2$ indicate that the system has the low dimensional nature and the antiferromagnetic in-plane correlation of the electrons in the region of $x$ < 0.6, as reported by the present author's group.[13]

The electronic nature of $\gamma$-Na$_x$CoO$_2$ has a rather sharp discontinuity at $x$ = 0.5,[17,18] that is, two successive phase transitions take place at temperatures, $T_{c1}$ = 87 K and $T_{c2}$ = 53 K, in the very narrow region of $x$ around 0.5. Recently studies of the transitions have revealed that the long-range antiferromagnetic order exists below $T_{c1}$ and a metal-insulator transition appears at $T_{c2}$ with decreasing $T$. However, on the mechanism(s) of the metal-insulator transitions, there does not exist firm understandings.[13,19,20]

The neutron results described above might indicate that the type of the stacking sequence of the CoO$_2$ layers, which affects the interlayer coupling, is important for the determination of the physical properties of Na$_{0.5}$CoO$_2$. (The stacking sequences of CoO$_2$ layers are schematically shown for $\gamma$- and $\beta$-phases in Fig. 1.) Then, in order to examine whether the stacking type affects the transitions, we have synthesized $\beta$-Na$_{0.5}$CoO$_2$ and carried out $^{59}$Co-NMR/NQR measurements. In addition, we have succeeded in preparing $\gamma$-K$_{0.5}$CoO$_2$ and found successive phase transitions similar to those of Na$_{0.5}$CoO$_2$. We have also carried out $^{59}$Co-NMR/NQR and neutron diffraction studies of $\gamma$-K$_{0.5}$CoO$_2$, which presents us useful data to be compared with those of $\gamma$-Na$_{0.5}$CoO$_2$.

## 2. Experimental details

Polycrystalline samples of $\beta$-Na$_{0.5}$CoO$_2$ were synthesized by the oxidation procedure from $\alpha$-NaCoO$_2$, which was prepared by means of the conventional solid-state reaction: The powder of Co metal and anhydrous NaOH pellets were mixed in a molar ration of Na : Co = 1.25 : 1 and heated in air at 500 ºC for 12 h, where we had to use the 25 % excess amounts of NaOH to obtain single phase of α-NaCoO$_2$. The fused ingots were finely ground and annealed at 800 ºC under the N$_2$ flow for 16 h. The obtained powder was immersed in the I$_2$/CH$_3$CN solution for 5 days to de-intercalate Na atoms and then the product was washed with CH$_3$CN and dried in air. We tried several different concentration of I$_2$ for the I$_2$/CH$_3$CN solution. After these processes, we obtained the samples of β-Na$_x$CoO$_2$ ($x \sim 0.5$), all of which were confirmed to be the single phase by powder X-ray diffraction studies.

*corresponding author : (e-mail : e43247a@nucc.cc.nagoya-u.ac.jp)

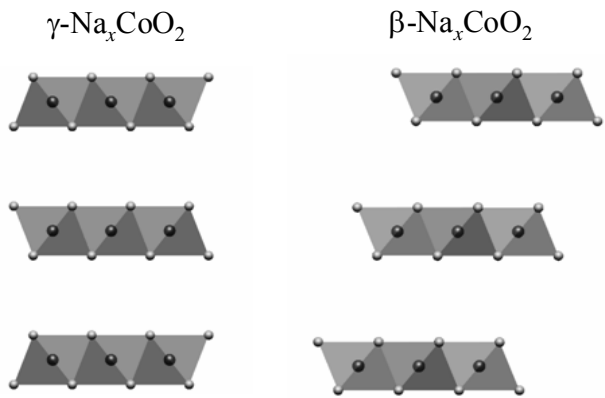

Fig. 1. Arrangements of CoO$_2$ planes in γ-Na$_{0.5}$CoO$_2$ (left) and β-Na$_{0.5}$CoO$_2$ (right), where Co sites represented by the dark spheres are within the shaded octahedra formed of the corner oxygens.

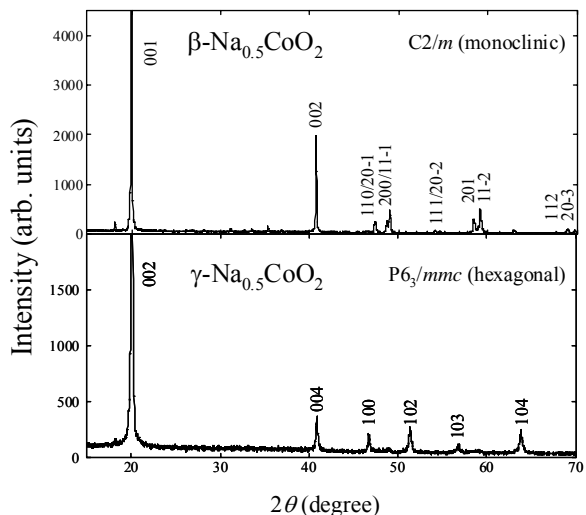

Fig. 2. X-ray diffraction patterns of β-Na$_{0.5}$CoO$_2$ (top) and γ-Na$_{0.5}$CoO$_2$ (bottom).

Single crystals of γ-K$_{0.5}$CoO$_2$ were synthesized by a floating zone method. A KOH and Co$_3$O$_4$ were mixed in the molar ration of K : Co = 1 : 1, where excess KOH was necessary for the complete reaction of Co$_3$O$_4$. Then, the obtained samples were heated in air at 400 ºC for 12 h. Finally, in order to remove the impurity phase K$_2$CO$_3$, the products were washed with C$_2$H$_5$OH. All peaks observed by the X-ray diffraction at room temperature could be indexed by the structure of γ-K$_{0.5}$CoO$_2$. (However, as is found later, peaks of the impurity phase K$_2$CO$_3$ were detected in the neutron diffraction data, which is because that the cross section of O atoms of K$_2$CO$_3$ is large for neutrons.) The lattice parameter $c$ was estimated to be ~12.453 Å at room temperature, which is significantly larger than the value of γ-Na$_{0.5}$CoO$_2$, ~11.129 Å.[17]

The electrical resistivity ρ was measured by a standard four-probe method in the temperature range from 4.2 to 300 K. Magnetization was measured in the magnetic field $H = 1$ T using a SQUID magnetometer (Quantum Design) in the temperature range from 2 to 350 K.

[59]Co-NMR and NQR measurements were performed for the randomly oriented powder samples by the spin-echo technique with a phase-coherent pulse spectrometer. The echoes were obtained using π/2-τ-π pulse sequences. The nuclear longitudinal relaxation rates $1/T_1$ of $^{59}$Co nuclei were estimated by measuring the $^{59}$Co nuclear magnetization $m$ as a function of the time $t$ elapsed after applying the inversion pulse in the frequency region around $3\nu_Q$, $\nu_Q$ being the electrical quadrupolar frequency. The $\{1-m(t)/m(\infty)\}$-$t$ curves were found to be described by the theoretical one.

The structural studies were carried out on ground samples by using a Rigaku X-ray diffractometer with FeKα radiation. The diffraction intensities were measured in the 2θ-range from 10˚ to 100˚ with a step of 0.02˚. In the temperature region higher than room temperature, a furnace was used and the temperature was controlled within an accuracy of 1 K. Rietveld analyses were carried out for these data by using Rietan2000.[21]

Powder neutron diffraction measurements were carried out for γ-K$_{0.5}$CoO$_2$ at 423 and 573 K using the high-resolution powder diffractometer (HRPD) installed at JRR-3 of JAERI in Tokai. The 331 reflections of Ge monochromator were used. The horizontal collimations were open (35')-20'-(sample)-6' and the neutron wavelength λ was ~1.823 Å. The diffraction intensities were measured in the 2θ-range from 2.5˚ to 165˚ with a step of 0.05˚. The temperature was controlled by a furnace. Rietveld analyses were carried out for these data by using Rietan2000.

## 3. Results and Discussion

### 3-1. β-Na$_{0.5}$CoO$_2$

In Figure 2, the observed X-ray diffraction patterns of β- and γ-Na$_{0.5}$CoO$_2$ are shown. Attached indices are referred to the space groups of C2/$m$ and P6$_3$/mmc for the former and latter systems, respectively.[22] In γ-Na$_{0.5}$CoO$_2$, Na atoms order with an orthorhombic superstructure commensurate with the Co lattice at 470 K.[18] and two distinct sites of Co$^{3.5+\delta}$ and Co$^{3.5-\delta}$ with different NMR quadrupolar frequencies are induced.[13] Presumably, the same type of Na ordering exists in β-Na$_{0.5}$CoO$_2$, but we have no definite information on this point. The lattice parameters for β-Na$_{0.5}$CoO$_2$ are $a = 4.891 (4)$, $b = 2.805(1)$, $c = 5.777(3)$ Å and β $= 105.77(8)$˚ and those for γ-Na$_{0.5}$CoO$_2$ are consistent with the ones reported previously.[17] The sodium content $x$ of β-Na$_x$CoO$_2$ was estimated by using the data shown in Fig. 3, where the distance between the CoO$_2$ layers are shown against the nominal sodium content $x$.[23-25]

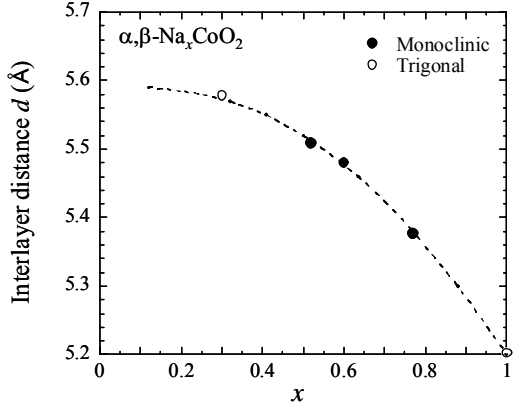

Fig. 3. Interlayer distance in α-, β-$Na_xCoO_2$ is plotted against $x$, where filled and open circles denote the monoclinic and trigonal unit cell, respectively. The dotted line is the guide for eyes.

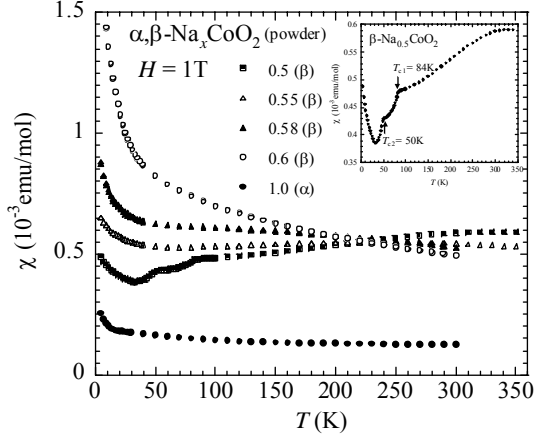

Fig. 4. Magnetic susceptibilities χ measured under $H$ = 1 T for various values of $x$ are plotted against $T$. Filled circles are the data of α-$NaCoO_2$ and the other four symbols correspond to β-phase samples. The data of β-$Na_{0.5}CoO_2$ are shown in the inset, too.

Although the $x$ value determined from this curve might have a certain error bar, the existence of the successive phase transitions similar to those of γ-$Na_{0.5}CoO_2$ suggests that β-phase samples with $x$ = 0.5 is successfully obtained (see the inset of Figure 4). The transition temperatures of the β-$Na_{0.5}CoO_2$ sample are $T_{c1}$ = 84 K and $T_{c2}$ = 50 K, which are about 3 K lower than those of γ-$Na_{0.5}CoO_2$. The effective interactions in γ-$Na_{0.5}CoO_2$ and β-$Na_{0.5}CoO_2$ seem to be very similar. Figure 4 shows the temperature dependence of the magnetic susceptibility χ measured with the magnetic field $H$ = 1 T for the powder samples of β-$Na_xCoO_2$ ($x$ ~ 0.5) and α-$NaCoO_2$. As $x$ decreases, the systematic change of the χ-$T$ curve from Curie-Weiss like to pseudogap-like behavior can be observed, which is similar to that reported previously for γ-$Na_xCoO_2$.[13] We note here that the absolute value of α-$NaCoO_2$ can be considered to correspond to the value of Van Vleck orbital susceptibility with low-spin nonmagnetic $Co^{3+}$.

From the macroscopic measurements presented above, we have not found significant differences in the magnetic behaviors between γ-$Na_{0.5}CoO_2$ and β-$Na_{0.5}CoO_2$. Then, we have performed $^{59}$Co-NMR/NQR, which can distinguish crystallographically distinct Co sites.

Field swept $^{59}$Co NMR spectra of β-$Na_{0.5}CoO_2$ taken at 90, 82, and 30 K with the frequency $f$ = 39.359 MHz are shown in Fig. 5. The obtained spectrum at 90 K (> $T_{c1}$) manifests the characteristics of the powder pattern of systems with anisotropic Knight shifts $K$ and the second order quadrupolar interaction. With decreasing $T$ through $T_{c1}$, the intensity of the central transition line exhibits significant decrease and its width increases symmetrically, indicating the growth of the long-range antiferromagnetic order. (The central line corresponds to the transition $I_z$ = 1/2 ↔ −1/2, where the $z$-axis is chosen that $V_{zz}$ has the maximum value among the eigen values of the electric-field-gradient (EFG) tensor, $V_{xx}$, $V_{yy}$ and $V_{zz}$.) In the presence of the internal magnetic field $H_{int}$, the central line is expected to locate at the frequency $f$ = $γH_{int}(1+K)$, where γ is the nuclear gyromagnetic ratio (γ = 10.03 MHz/T for $^{59}$Co nuclei).

NMR/NQR spectra taken at $T$ = 90, 60, and 10 K in the zero external magnetic field are shown in Fig. 6 with the $f^{2}$-correction. The data of β-$Na_{0.5}CoO_2$ are shown in the inset, too. The calculated pattern at 10 K is also shown in the figure. Two resonance lines observed at 90 K (> $T_{c1}$) (top panel) are the third satellites of the NQR signals from two distinct Co sites with different $ν_Q$ values, $ν_Q^1$ ~ 4.0 MHz and $ν_Q^2$ ~ 2.9 MHz. (The superscripts 1 and 2 indicate the sites Co1 and Co2.) In the data at 90 K, we have not found any evidence for the magnetic order. Hereafter, as shown in Fig. 6, we attach circles or triangles to the signals, to distinguish the two Co sites with the smaller or larger $ν_Q$ values, respectively.

Below $T_{c1}$, two sets of peaks composed of one center peak and three pairs of satellite peaks were observed. (In the figure, center peaks of two sets are indicated by the filled circle and filled triangle.) The peaks indicated by circles split into two below $T_{c2}$, which was also observed for γ-$Na_{0.5}CoO_2$.[13,20] Although this behavior might be related with the metal-insulator transition at $T_{c2}$, we do not discuss it here.

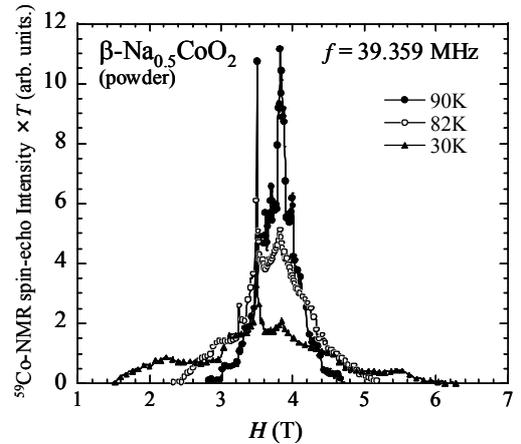

Fig. 5. Field-swept $^{59}$Co-NMR spectra of β-$Na_{0.5}CoO_2$ taken at $T$ = 30, 82 and 90 K with a fixed frequency of 39.359 MHz for randomly oriented powder samples.

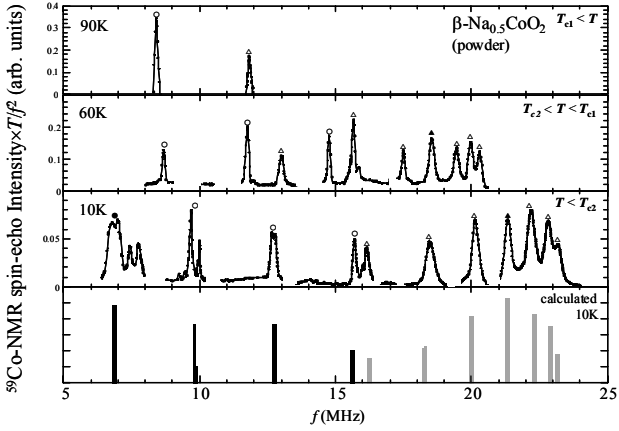

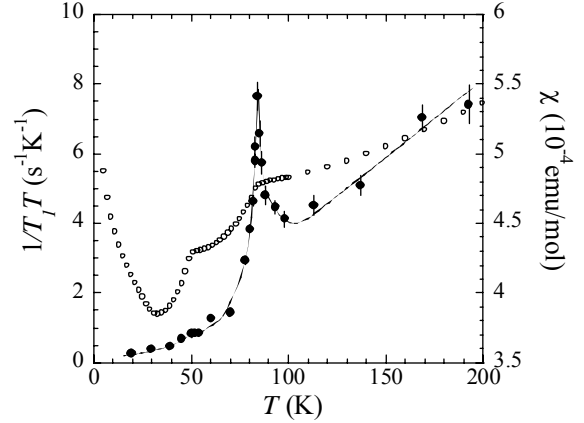

Fig. 6. Zero-field NMR spectra taken at 90, 60 and 10 K. The circular and triangular symbols show the signals from different Co sites with $\nu_Q \sim$ 2.8 MHz and 4 MHz, respectively. The filled symbols denote the center line ($I_z = 1/2 \leftrightarrow -1/2$) of each Co site. The bottom panel shows the calculated spectra at 10 K.

Fig. 7. $T$-dependence of the $1/T_1T$ for the Co sites with $\nu_Q^1 \sim$ 4.0 MHz (filled circles) is shown together with the uniform magnetic susceptibility (open circles).

To find the magnitude and direction of $\boldsymbol{H}_{\mathrm{int}}$, we tried to fit calculated results to the spectra at 10 K ($< T_{c2}$). In this calculation, the electrical quadrupole interaction has been treated as a perturbation up to the second order. The calculated peak positions are shown in the bottom panel of Fig. 6. The NMR parameters obtained are as follows: The quadrupolar frequencies $\nu_Q^1 = 3.95$ MHz and $\nu_Q^2 = 2.95$ MHz, the internal magnetic fields $\boldsymbol{H}_{\mathrm{int}}^1 = 1.98$ T (// $ab$-plane) and $\boldsymbol{H}_{\mathrm{int}}^2 = 0.66$ T (// $c$), and the asymmetric parameters $\eta^1 = \eta^2 = 0.44$. They are rather similar to those obtained for $\gamma$-phase except for $\eta$.[13] The estimated values of $\eta$ for the present $\beta$-Na$_{0.5}$CoO$_2$ is about twice as large as that of $\gamma$-Na$_{0.5}$CoO$_2$. It is likely that the monoclinic distortion of the unit cell may cause this large difference of the $\eta$ values. However, for the satisfactory explanation of the difference, the structure of $\beta$-Na$_{0.5}$CoO$_2$ has to be studied in detail, as has been done for $\gamma$-Na$_{0.5}$CoO$_2$.[17]

In Fig. 7, the longitudinal relaxation rate divided by temperature, $1/T_1T$ measured at Co sites with $\nu_Q^1 \sim 4.0$ MHz is plotted against $T$ together with the data of magnetic susceptibility $\chi$ shown in Fig. 3. As one can see from the figure, the $1/T_1T$-$T$ curve shows a prominent peak and a small kink at temperatures corresponding to the anomalies in $\chi$.

Now, both $\beta$- and $\gamma$-Na$_{0.5}$CoO$_2$ systems have been found to have, in ordered state, almost same transition temperatures and similar NMR parameters except $\eta$. Very similar behaviors of $\chi$ and $1/T_1T$ are also observed. At least, the difference between the stacking sequences of these two systems does not bring about significant change of the physical properties. The result is consistent with the results of the studies on the superconducting three-layer system reported by the present authors' group.[26]

According to ref. 27, the origin of interlayer coupling in Na$_x$CoO$_2$ can be understood by the superexchange interaction via the direct O-O hopping and through the intermediate Na $sp^2$ hybridized orbits, where each Co spin has the interlayer coupling to the spins of the nearest and the six next-nearest neighbor ones of the adjacent CoO$_2$ planes. This multiple exchange paths induce the three-dimensional-like nature of the dispersion of the magnetic excitations in the ordered state of $\gamma$-Na$_x$CoO$_2$ for $x > 0.7$, but the individual exchange interaction has rather two-dimensional nature, like the lattice itself. For all of these reasons, roughly speaking, the stacking sequence of CoO$_2$ layers might not affect the interlayer interaction.

### 3-2. $\gamma$-K$_{0.5}$CoO$_2$

In Figure 8, the electrical resistivities $\rho$ and magnetic susceptibilities $\chi$ for an as-grown sample of $\gamma$-K$_{0.5}$CoO$_2$ and a sample of $\gamma$-Na$_{0.5}$CoO$_2$ synthesized by the oxidation procedure from $\gamma$-Na$_{0.72}$CoO$_2$[13], are compared, where $\gamma$-K$_{0.5}$CoO$_2$ is found to have similar successive phase transitions to those of γ-$Na_{0.5}CoO_2$. However, the transition temperatures in γ-$K_{0.5}CoO_2$ are about $T_{c1}$ = 60 K and $T_{c2}$ = 20 K, which are about 30 degrees lower than those of γ-$Na_{0.5}CoO_2$. For both systems, the significant suppression of χ is observed only for *H* // *ab*-plane. It can be understood by considering that the direction of the larger moments is within the *ab* palne, as was discussed in our previous paper for γ-$Na_{0.5}CoO_2$.[13] The NMR data shown later for γ-$K_{0.5}CoO_2$ are consistent with the idea. With further decreasing $T$, the magnetic susceptibilities χ begin to decrease at $T_{c2}$ for both magnetic field directions of *H* // *ab*-plane and *H* // *c*. The resistivities ρ exhibit only a tiny anomaly at $T_{c1}$ and exhibit a sharp increase at $T_{c2}$ with decreasing $T$. There may exist another anomaly in the slope of the ρ-$T$ curve below $T_{c2}$, indicating another phase transition or the non-negligible effects of the electron localization.[17] It should be also noted that the ρ value of γ-$Na_{0.5}CoO_2$ is about five times larger than that of γ-$K_{0.5}CoO_2$. The difference between the sample preparation processes may cause this difference.

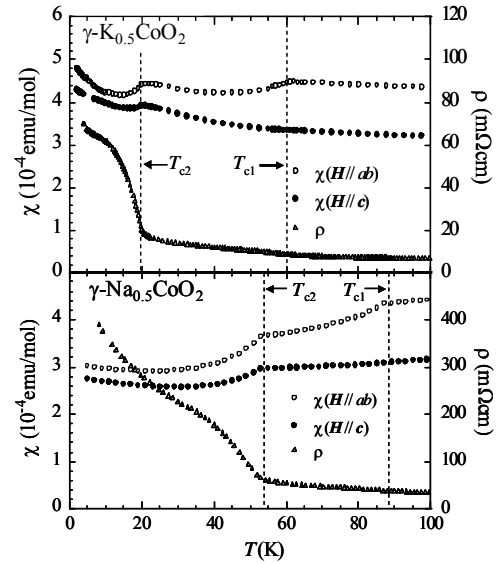

Fig. 8. $T$-dependence of the magnetic susceptibilities $\chi$ measured under $H = 1$ T parallel to $ab$-plane (open circles) & $c$-axis (filled circles) and electrical resistivities $\rho$ (open triangles) for a single crystal of $\gamma$-K$_{0.5}$CoO$_2$ (top panel) and $\gamma$-Na$_{0.5}$CoO$_2$ (bottom panel). $T_{c1}$ and $T_{c2}$ are 60 K and 20 K for the former and 87 K and 53 K for the latter, respectively.

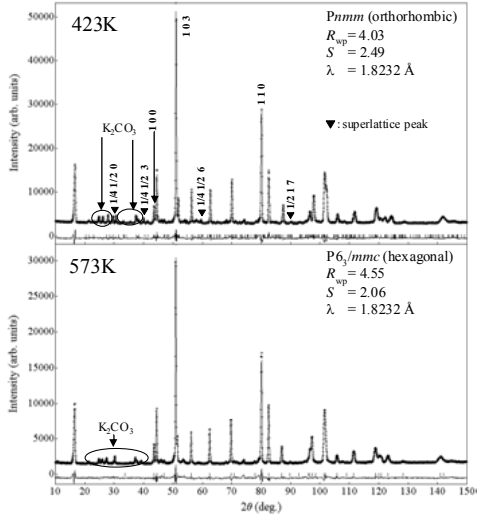
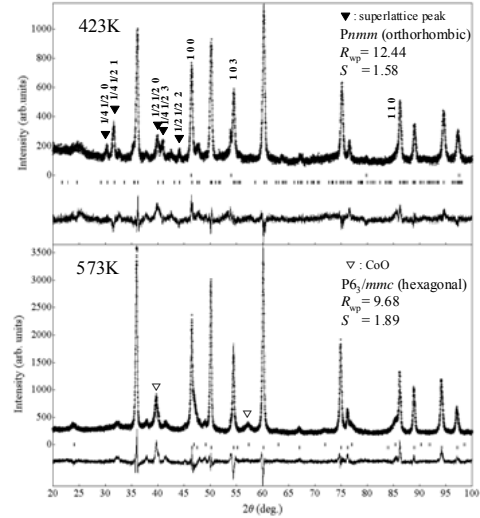

Fig. 9. Neutron powder diffraction patterns of γ-$K_{0.5}CoO_2$ at 423 (top panel) and 573 K (bottom panel). To several diffraction peaks, the indices referred to the high temperature hexagonal phase (space group P$6_3$/mmc) are attached. Solid Triangles indicate the superlattice peaks. The results of the fittings carried out by fixing the K atoms at the positions determined by the X-ray Rietveld analyses at the corresponding temperatures, are also shown by the solid lines. See text for details.

Fig. 10. X-ray powder diffraction patterns of γ-$K_{0.5}CoO_2$ at 423 (top panel) and 573 K (bottom panel). To several diffraction peaks, the indices referred to the high temperature hexagonal phase (space group P$6_3$/mmc) are attached. Solid Triangles indicate the superlattice peaks. The results of the fittings carried out by fixing the O atoms at the positions determined by the neutron Rietveld analyses at the corresponding temperatures, are also shown by the solid lines.

We have shown that γ-$K_{0.5}CoO_2$ has the phase transitions similar to those of γ-$Na_{0.5}CoO_2$. Then, to see the detailed difference of the observed data between γ-$K_{0.5}CoO_2$ and γ-$Na_{0.5}CoO_2$, we have carried out structural studies of γ-$K_{0.5}CoO_2$.

Figure 9 shows the observed neutron diffraction data of γ-$K_{0.5}CoO_2$ at 423 and 573 K. A number of superlattice peaks arising from the potassium order can be observed at 423 K, as indicated by the filled triangles. These peaks can be indexed by using the orthorhombic space group P$nmm$, as for γ-$Na_{0.5}CoO_2$. Then, the temperature was raised to 573 K, and the diffraction patterns were taken. At the temperature, the superlattice peaks observed at 423 K disappear and all the peaks are indexed by the hexagonal space group P$6_3$/mmc except those from the impurity phase of $K_2CO_3$. Then, calculating the diffraction curve for the mixtures of γ-$K_{0.5}CoO_2$ and $K_2CO_3$, we have carried out the Rietveld analyses. At both temperatures of 423 and 573 K, the error bars of the positional parameters of K atoms (at 6h sites of the space group P$6_3$/mmc at 573 K and 2a and 2b sites of the space group P$nmm$ at 473 K) obtained by the fitting were rather large, though the positions of O atoms can be determined. It is because the neutron scattering amplitude of K atoms is very small as compared with that of O atoms. (Here, it is noted that the large error bars for the K atoms do not have a significant effect on the positions of Co. It is also noted that the Co atoms do not have the positional parameters for the present space group.)

To determine the precise K atom positions, we carried out powder X-ray diffraction studies for another sample of γ-$K_{0.5}CoO_2$, too. The results are shown at 423 and 573 K in Fig. 10. At 423 K, we could observe the superlattice peaks more clearly than in the neutron measurements. (The superlattice peaks are indicated by the filled triangles.) After the measurements, the temperature was raised to 573 K, the superlattice peaks disappear and the peaks of the CoO impurity phase were observed as indicated by the open triangles. It suggests that γ-$K_{0.5}CoO_2$ tends to decompose at the elevated temperature. Here, considering the contribution of the CoO impurity phase, we have carried out the Rietveld analyses in the 2θ-range from 20° to 100° by fixing the positional parameters of O atoms at the values determined by the neutron Rietveld analyses, which enables us to determine the K atom positions

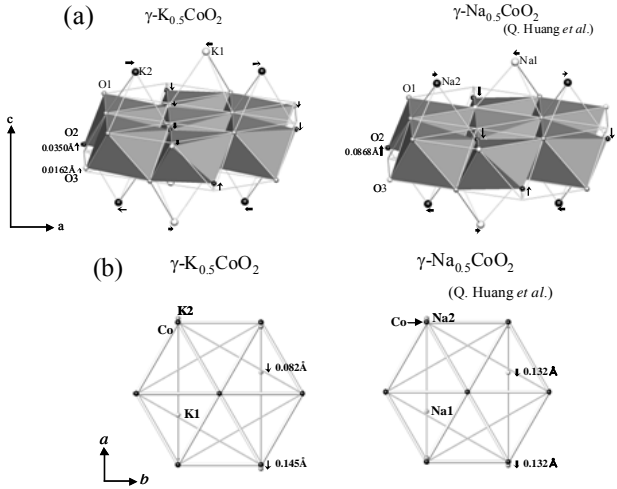

Fig. 11. (a) Structures of γ-$K_{0.5}CoO_2$ determined by the present Rietveld analyses is shown schematically and compared with that of γ-$Na_{0.5}CoO_2$.[17] The arrows indicate the atomic shift from the position corresponding to the hexagonal symmetry. The shifts of oxygen atoms along *c*-axis are five times as large as the actual values. (b) Arrangements of Co and K or Na atoms are shown, by projecting them onto the *c* plane. The arrows show the atomic shifts for potassium (sodium) atoms within the *ab*-plane from the positions above the cobalt atoms or from that above the center of gravity of the triangle formed by three Co atoms.

more precisely than the case of neutron studies.

Then, we have carried out the Rietveld analyses of the neutron data, again, by fixing the K atom positions at the values determined by the X-ray Rietveld analyses stated above. The results are shown in Fig. 9 by the solid lines. The final structural parameters at 423 and 573 K are shown in Table 1.

In Figs. 11(a) and 11(b), the structure of γ-$K_{0.5}CoO_2$ determined by the Rietveld analyses is shown schematically and compared with that of γ-$Na_{0.5}CoO_2$,[17] where K1 (Na1), K2 (Na2) and O1, O2, O3 represent crystallographically distinct sites. K1 (Na1) atoms are on the line along *c* through the center of gravity of a triangle formed by three Co atoms within an *ab*-plane and K2 (Na2) atoms lie above the Co atoms (or on the same line along *c* through the Co atoms). The arrows indicate the atomic shifts from the positions corresponding to the hexagonal symmetry described by thin bonds. In the figure, because the atomic shifts of oxygen atoms along *c*-axis are very small as compared to those of sodium and potassium atoms, they are drawn five times longer than the actual ones. The degrees of the structural distortion of the $CoO_6$ octahedra from the hexagonal symmetry are different between γ-$K_{0.5}CoO_2$ and γ-$Na_{0.5}CoO_2$.

In order to determine the transition temperature $T_{c0}$ of the ordering of K atoms by X-ray powder diffraction, the *T*-dependence of the integrated intensities of the 1/4 1/2 1 reflection has been measured and the results are in Fig. 12. We have not detected appreciable hysteretic effects. We can see that the second order transition takes place at $T_{c0} \sim 550$ K, which is about 80 K higher than that of γ-$Na_{0.5}CoO_2$.[18] Perhaps this difference between the $T_{c0}$ values is attributed to the mass difference between the ordered atoms.

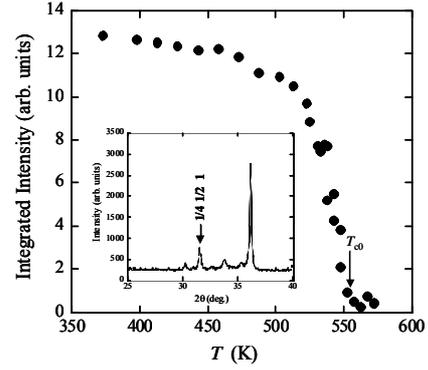

Fig. 12. *T*-dependence of the integrated intensities of the 1/4 1/2 1 superlattice reflection in the temperature region from 373 K to 573 K. Inset shows the X-ray powder diffraction pattern of γ-$K_{0.5}CoO_2$ around the 1/4 1/2 1 reflection.

Figure 13 shows the NQR spectra of γ-$K_{0.5}CoO_2$ at 78 K above $T_{c1}$, where data of γ-$Na_{0.5}CoO_2$ at 90 K are also shown for comparison. NQR peaks were observed at the positions corresponding to the frequencies of $3\nu_Q$ for the two distinct Co sites with $\nu_Q^1 \sim 3.80(4.01)$ MHz and $\nu_Q^2 \sim 3.07(2.77)$ MHz in γ-$K_{0.5}CoO_2$ (γ-$Na_{0.5}CoO_2$). The absolute value of $\nu_Q^1-\nu_Q^2$ of γ-$K_{0.5}CoO_2$ is smaller than that of γ-$Na_{0.5}CoO_2$, that is, the difference between the Co-valences or the EFG's at Co sites of the former is smaller than that of the latter. Then, what is the main origin of this difference of $\nu_Q$? Calculations by using the point charge model for the structural parameters discussed above indicate that the observed structural difference between the two does not induce a significant difference in the EFG, suggesting that the difference between the Co-valences is the main cause of the observed difference. On this point, further information can be found from the zero-field NMR data described below.

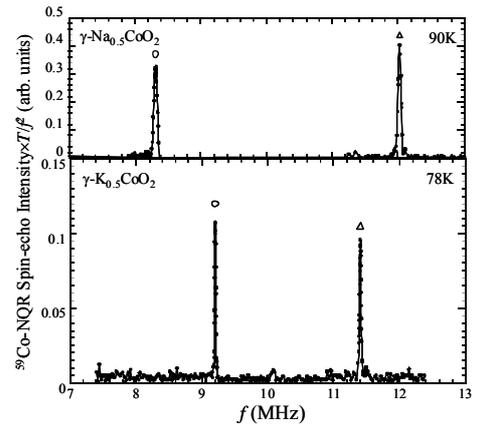

Fig. 13. NQR spectra of γ-$Na_{0.5}CoO_2$ (top panel) and γ-$K_{0.5}CoO_2$ (bottom panel) above $T_{c1}$. NQR peaks corresponding to the frequencies of $3\nu_Q$ for the two distinct Co site with $\nu_Q^1 \sim 3.80$ (4.01) MHz and $\nu_Q^2 \sim 3.07$ (2.77) MHz in γ-$K_{0.5}CoO_2$ (γ-$Na_{0.5}CoO_2$), were observed as indicated by the triangles and circles.

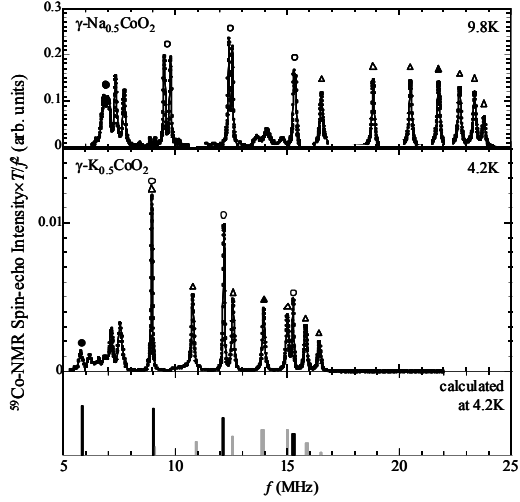

Fig. 14. Zero-field NMR spectra for γ-$K_{0.5}CoO_2$ and γ-$Na_{0.5}CoO_2$ below $T_{c2}$. The circular and triangular symbols represent the signals from the two distinct sites with different values of $\nu_Q$. The filled symbols denote the center line ($I_z = 1/2 \leftrightarrow -1/2$) of each Co site. The bottom panel shows the calculated spectra at 10 K.

In Fig. 14, the Zero field NMR spectra of γ-$K_{0.5}CoO_2$ are compared with those of γ-$Na_{0.5}CoO_2$. For the former, peaks indicated by the circles do not split into two below $T_{c2}$, which is in contrast with the result of the latter. As shown in the bottom panel, the spectra at 4.2 K can be explained by simple calculations with the following parameter sets: [$\nu_Q^1$ = 3.82 MHz, $\eta^1$ = 0.70, $H_{int}^1$ = 1.29 T (// $ab$-plane), $\nu_Q^2$ = 3.15 MHz, $\eta^2$ = 0.20 and $H_{int}^2$ = 0.57 T (// $c$)] and [$\nu_Q^1$ = 3.07 MHz, $\eta^1$ = 0.20, $H_{int}^1$ = 1.29 T (// $ab$-plane), $\nu_Q^2$ = 3.15 MHz, $\eta^2$ = 0.20 and $H_{int}^2$ = 0.57 T (// $c$)]. The result that the larger internal field is within the $ab$-plane, is consistent with that stated above from the data of the magnetic susceptibility χ. For the former parameter set, an unsettled question remains whether such a large $\eta^1$ value is reasonable or not. For the latter set, the $\nu_Q^1$ value of 3.07 MHz does not agree with the one estimated from the NQR spectra at $T$ = 78 K. Because it is quite unlikely that only $\nu_Q^1$ has the temperature dependence, in contrast to the $T$-independent behavior of $\nu_Q^1$ and $\nu_Q^2$ found for γ- and β-$Na_{0.5}CoO_2$, the former set of parameters is more plausible than the latter.

We have shown that γ-$K_{0.5}CoO_2$ system exhibits qualitatively similar physical behaviors to those of γ-$Na_{0.5}CoO_2$. For example, the behaviors in χ and ρ, the order pattern of alkali ions and the direction of internal magnetic fields are very similar, though absolute values of $T_{c1}$, $T_{c2}$ and $T_{c0}$ and NMR parameters in magnetically ordered state are different from those of $Na_{0.5}CoO_2$.

| Atom | site | $x$ | $y$ | $z$ | Occu. | $B$(Å$^2$) |
|---|---|---|---|---|---|---|
| Co1 | 4f | 0 | 1/4 | 0 | 1 | 0.64(5) |
| Co2 | 4d | 1/2 | 0 | 0 | 1 | 0.64(5) |
| K1 | 2b | 0.970(2) | 1/4 | 3/4 | 1 | 1.80(6) |
| K2 | 2a | 0.350(2) | 3/4 | 3/4 | 1 | 1.80(6) |
| O1 | 4f | 1/3 | 1/4 | 0.0772(6) | 1 | 0.58(2) |
| O2 | 4f | 1/3 | 3/4 | 0.0757(6) | 1 | 0.58(2) |
| O3 | 8g | 5/6 | 0 | 0.0785(3) | 1 | 0.58(2) |

Space group P$nmm$; $a$ = 4.9068(4) Å, $b$ = 5.6627(5) Å, $c$ = 12.5028(4) Å, $R_{wp}$ = 4.03, $S$ = 2.49

| Atom | site | $x$ | $y$ | $z$ | Occu. | $B$(Å$^2$) |
|---|---|---|---|---|---|---|
| Co | 2a | 0 | 0 | 0 | 1 | 1.07(6) |
| K1 | 6h | 0.089(2) | 0.1783(4) | 1/4 | 0.065(3) | 1.90(5) |
| K2 | 6h | 0.755(2) | 0.511(4) | 1/4 | 0.100(3) | 1.90(5) |
| O | 4f | 1/3 | 2/3 | 0.07768(8) | 1 | 0.78(2) |

Space group P6$_3$/$mmc$; $a$ = 2.8330(1) Å, $c$ = 12.5812(4) Å, $R_{wp}$ = 4.55, $S$ = 2.05

Table 1. Atomic coordinate of γ-$K_{0.5}CoO_2$ at 423 (top) and 573 K (bottom).

## 4. Summary

We have shown that β-$Na_{0.5}CoO_2$ and γ-$K_{0.5}CoO_2$ exhibit the successive phase transitions at $T_{c1}$ and $T_{c2}$, which are very similar to those of γ-$Na_{0.5}CoO_2$. The transition temperatures $T_{c1}$ and $T_{c2}$ are 84 and 50 K for β-$Na_{0.5}CoO_2$ and 60 and 20 K for γ-$K_{0.5}CoO_2$, respectively. $^{59}$Co-NMR measurements have revealed that the internal magnetic field due to the antiferromagnetic order appears at $T_{c1}$ with decreasing $T$ in both materials. Similar results for γ-$Na_{0.5}CoO_2$ were already reported by the present authors' group. Both systems of β- and γ-$Na_{0.5}CoO_2$ have almost same $T_{c1}$, $T_{c2}$ and NMR parameters below $T_{c2}$. $1/T_1T$ and $\chi$ of these two systems exhibit very similar behavior, too. These results indicate that the stacking sequences of $CoO_2$ layers between β- and γ-$Na_{0.5}CoO_2$ seem not to induce significant differences between the physical properties of these systems. Even for γ-$K_{0.5}CoO_2$, various physical behaviors observed by the NMR studies are qualitatively similar to those of β- and γ-$Na_{0.5}CoO_2$, though $T_{c1}$ and $T_{c2}$ are smaller than those of β- and γ-$Na_{0.5}CoO_2$ by about 30 K. This difference may be explained by the difference between the distances of $CoO_2$ layers.

## Acknowledgements

This work is supported by Grants-in-Aid for Scientific Research from the Japan Society for the Promotion of Science (JSPS) and by Grants-in-aid on priority areas from the Ministry of Education, Culture, Sports, Science and Technology. Work at JRR-3M was performed within the frame of JAERI Collaborative Research Program on Neutron Scattering.